\documentclass[graybox]{svmult}

\usepackage{mathptmx}       %
\usepackage{amssymb}
\usepackage{helvet}         %
\usepackage{courier}        %
\usepackage{type1cm}        %
\usepackage{makeidx}         %
\usepackage{graphicx}
\usepackage{amsmath}
\usepackage{enumitem} %
\graphicspath{{Figures/}}
\usepackage{multicol}        %
\usepackage[bottom]{footmisc}%

\makeindex             %

\begin{document}

\title*{Weighted temporal event graphs}
\author{Jari Saram\"aki and Mikko Kivel\"a and M\'arton Karsai}
\institute{Jari Saram\"aki \at Department of Computer Science, Aalto University School of
  Science, P.O.  Box 15400,  FI-00076, Finland, \email{jari.saramaki@aalto.fi}
\and Mikko Kivel\"a \at Department of Computer Science, Aalto University School of
  Science, P.O.  Box 15400,  FI-00076, Finland, \email{mikko.kivela@aalto.fi}
  \and M\'arton Karsai \at Univ. Lyon, ENS de Lyon, Inria, CNRS, UCB Lyon 1, LIP UMR 5668, IXXI, F-69342, Lyon, France, \email{marton.karsai@ens-lyon.fr}}
\maketitle

\abstract{The times of temporal-network events and their correlations contain information on the function of the network and they influence dynamical processes taking place on it. To extract information out of correlated event times, techniques such as the analysis of temporal motifs  have been developed. We discuss a recently-introduced, more general framework that maps temporal-network structure into static graphs while retaining information on time-respecting paths and the time differences between their consequent events. This framework builds on weighted temporal event graphs: directed, acyclic graphs (DAGs) that contain a superposition of all temporal paths. We introduce the reader to the temporal event-graph mapping and associated computational methods and illustrate its use by applying the framework to temporal-network percolation.}

\section{Introduction}

There are two key reasons behind the success of the temporal networks framework~\cite{temporalnetworks,Holme_modern}. Both have to do with the rich additional information brought by knowing  the specific times of interactions between nodes. First, the times of interaction events and their correlations contain detailed information about the dynamics of the entities that form the network. Consider, as an example,  studies in computational social science that build on data of human communication: the time stamps of communication events carry far more information on human behaviour than any static network mapping would~(see, e.g.,~\cite{Karsai2011,jo2012,miritello_limited_2013,aledavood_daily_2015,Navarro2017}). Second, the times of events and their correlations can strongly influence dynamical processes taking place on networks. Their effect can be so strong that the static-network picture can become invalid for some dynamical processes \cite{Karsai2011,Iribarren2009,Horvath2014}. Therefore, more often than not, the times of interactions simply have to be taken into account if one wants to obtain an accurate understanding of the dynamics of processes that unfold on temporal networks.

For both of the above goals -- extracting information from the network itself and understanding how the network affects dynamical processes -- new kinds of mathematical and computational tools are required. While many static-network concepts can be extended to temporal networks (at least roughly), the additional degree of freedom due to the temporal dimension complicates things. Computing network measures for temporal networks often requires approaches that are very different from the static case. But even simply defining the measures properly may be less than straightforward. Consider, \emph{e.g.}, shortest paths between nodes: in a static, unweighted network, the only attribute of a shortest path (if it exists) is its length, and it is readily discovered by a breadth-first search. In temporal networks, when considering shortest paths, one has to define "short" first -- does it mean the fastest path, or the one with the smallest number of events, or maybe corresponding to the shortest static-network path? Then, additionally, one has to choose the time frame that one is interested in, as paths are fleeting entities that are only brought about by their constituent events: even if there is a path now, there may be none a second later. And then, finally, one has to devise a computational method for empirical data that extracts the shortest temporal paths in a reasonable amount of time. 

Nevertheless, it would be convenient to repurpose computational and theoretical methods developed for analyzing static networks for temporal-network studies, because there is an abundance of such methods. This would become possible, \emph{e.g.}, if one was able to cast temporal networks as static entities so that the properties of those static entities correspond to the properties of the original temporal networks (though not necessarily in an one-to-one way). Generally speaking, this is not straightforward; some approaches have been introduced in the literature that typically focus on some chosen subset of the properties of  temporal networks (see, \emph{e.g.}, \cite{Nicosia2012}). 

In this Chapter, we discuss an approach that maps temporal-network structure onto a weighted static event graph~\cite{Kivela2018} that is directed and acyclic and whose weights encode the time differences between events. This mapping is done so that time-respecting paths of the original network are preserved. Temporal-network event graphs are analogous to line graphs of static networks~\cite{Mellor2017}. In the type of event graphs discussed here, nodes represent events of the original temporal network,  directed links connect events that share a node in the temporal network so that their direction follows the direction of time, and the link weights indicate the time difference between the two events that the link connects. As a concrete example, if A calls B who then calls C, the weighted event graph would have two nodes (the AB call and the BC call), so that the AB node is connected to the BC node with a directed link whose weight is the waiting time from the AB call to the BC call. 

As its main strength, this approach encodes temporal information as a static network structure. This then
allows %
extraction of temporal-network structures that are constrained by the time differences $\Delta t$ between successive events, from temporal motifs to time-respecting paths whose events have to follow one another within some time limit and to  temporal components defined by connectivity through such time-respecting paths.
Additionally, one can use known static-graph-based methods and find these structures
in a way that is computationally extremely efficient as compared to brute-force methods applied to the original temporal network.
 For example, one can use the method developed for percolation studies \cite{Newman2001Fast} %
where one performs sweeps of activating one connection at a time, in this case in the order of increasing time difference  $\Delta t$. 
This method is computationally much more efficient compared to brute-force breadth-first-search approaches, which have been used in temporal-network studies. Such approaches were also used in  conventional percolation studies before more efficient methods were discovered \cite{Leath1976Cluster}.

Being able to quickly obtain temporal-network paths and components and sweep through a range of constraints is particularly useful 
for studying spreading or transportation processes that have to leave a node within some time limit $\Delta t$. To mention a few, such processes include variants of the common models of contagion, such as Susceptible-Infectious-Recovered and Susceptible-Infectious-Susceptible, where the recovery/infectiousness time is assumed to be constant or has a clear upper bound. Other types of dynamics include social contagion where information or rumours age, ad-hoc message passing by mobile agents that keep the message only for a limited time, and routing of passengers in transport networks, where both lower and upper limits on the possible or acceptable transfer time may exist. 

Once the weighted event graph has been constructed from the temporal network, one can quickly extract subnetworks that correspond to chosen values of $\Delta t$; we shall discuss how this is done below. These subgraphs, being static, can then be approached using static-network methods and algorithms. There can be additional computational advantages because these subnetworks are directed and acyclic, and there are fast methods developed for directed acyclic graphs. Further, as discussed above, because the event graph encodes all time differences between events, one can quickly sweep through a range of differences to see how the maximum (or minimum) difference affects the outcome, for example the existence of temporally connected components.

This Chapter is structured as follows. We begin by 
providing definitions for concepts related to temporal adjacency and connectivity that are required for constructing the event graphs. We then continue by discussing how temporal networks can be mapped to weighted event graphs, both in theory and in practice. We next talk about how to interpret the structural properties of weighted event graphs: how their topological features (such as directed paths or weakly connected components) map back onto the original temporal networks. This discussion is followed by an examples of applications of this framework to temporal motif analysis and to temporal-network percolation studies. Finally, we present our conclusions and discuss possible future directions.

\section{Mapping temporal networks onto weighted event graphs}

\subsection{Definitions: vertices, events, temporal network}

Let us consider a temporal network $G=(V_G,E_G,T)$, where $V_G$ is the set of vertices and $E_G\subset V_G\times V_G \times \left[0,T\right]$ is the finite set of interaction events between the vertices with known times, so that the interactions take place within some limited period of observation $[0,T]$ that can also be considered to be the lifetime of $G$. We  denote an interaction event -- simply called an "event" from now on -- between vertices $i$ and $j$ at time $t$ with $e(i,j,t)$. Please note that in the following, we require that one node is only allowed to participate in one event at any given point in time.

Note that depending on the context, the events may be directed (\emph{e.g.}, representing an email from $i$ to $j$ in email data) or undirected (\emph{e.g.}, representing a face-to-face conversation between two persons). This choice has consequences on dynamical processes taking place on top of the temporal network: in the case of social contagion, for example, one email or one text message carries information one way only, while a face-to-face conversation can carry information both ways.

There are also cases where the events have a non-zero \emph{duration} that has to be taken into account in temporal-network studies. Examples include the flights in a passenger's route in an air transport network and phone calls in a communication network -- 
in both cases, a new event (the connecting flight, the next phone call) cannot begin before the first event is finished.
When the event duration needs to be taken into account, events are defined as quadruples, $e(i,j,t,\tau)$, where $\tau$ indicates the duration of the event.

Depending on the type of events in a temporal network, the time difference is defined in slightly different way:
\begin{definition}
\emph{Time difference between events.} Given two events $e_1$ and $e_2$, their time difference $\delta t (e_1,e_2)$ is either the difference in times $\delta t (e_1,e_2)=t_2 - t_1$ for \emph{events without duration} or the time from the end of the first event to the start of the second one $\delta t (e_1,e_2)=t_2 - t_1 - \tau_1$ for \emph{events with duration}. \label{def:deltat}
\end{definition}
Note that these two definitions become the same if the events have zero duration.

\subsection{Definitions: adjacency and $\Delta t$-adjacency}

Our goal is to investigate larger temporal-network structures, from mesoscopic to macroscopic entities, that arise out of the topological and temporal correlations of the network's constituent events. We begin by defining criteria for events being topologically and temporally close to one another and then move on to defining larger entities based on these criteria.
The following concepts of temporal adjacency, temporal connectivity, and temporal subgraphs are the building blocks for temporal-network event graphs as well as their substructures from components to temporal motifs. The concept of temporal adjacency also leads straighforwardly to the notion of time-respecting paths.

\begin{definition}\emph{Temporal adjacency.}
Two events $e_1(i,j,t_1)$ and $e_2(k,l,t_2)$ are \emph{temporally adjacent}, denoted $e_1\rightarrow e_2$, if they share (at least) a node, $|\{ i,j\} \cap \{k,l\}| >0$, and they are consecutive (but not simultaneous) in time, \emph{i.e.} 
 $\delta t(e_1,e_2)>0$.
  \label{def:temporal_adjacency}
\end{definition}

\begin{definition}\emph{$\Delta t$-adjacency.} Two  events $e_1$ and $e_2$ are \emph{$\Delta t$-adjacent}, denoted $e_1\xrightarrow{\Delta t} e_2$, if they are temporally adjacent and the time difference between them is $\delta t(e_1,e_2) %
\leq \Delta t$. \label{def:delta_t_adjacency}
\end{definition}

Temporal adjacency and $\Delta t$-adjacency are always directed regardless of whether the events themselves are directed or not, and their direction follows the direction of time, from the event that took place first to the event that took place next. Please note that here we use a directed definition of adjacency unlike in \cite{Kovanen_JSTAT2011,Kovanen_book,Kovanen_PNAS} for reasons that will become apparent later.

Depending on the problem at hand, one may wish to include additional constraints in the definition of temporal adjacency. If the original events are directed and this directionality is important, \emph{e.g.}, because it affects information flows, it can be directly incorporated into the definition of temporal adjacency, so that $e(i,j,t)$ and $e(j,k,t+1)$ are considered adjacent, while $e(i,j,t)$ and $e(k,j,t+1)$ are not (see Def.~\ref{def:temporal_adjacency}). This will also affect time-respecting paths defined according to Def.~\ref{def:time_respecting_path}. It is possible to introduce further constraints, \emph{e.g.}, ignoring return events (non-backtracking events only) which might be useful for modelling certain types of spreading dynamics. In this case, the pair $e(i,j,t)$ and $e(j,i,t+1)$ should not be added to $G$.

One can also consider allowing simultaneous interactions of the same node by introducing \emph{hyper-events} and an adjacency relationship between them for the definition of a hyper-event graph. In this extension, events happening at the same time and sharing nodes may be grouped in a hyper-event, which this way represents a set of simultaneous events as a single object. Two hyper-events taking place at times $t$ and $t'$ are adjacent if they are consecutive ($t<t'$) and share at least one node from the set of nodes they involve~\cite{karsaiUnpubl2019}. 

\subsection{Definitions: temporal connectivity and temporal subgraphs}

To study the mesoscopic building blocks of temporal structures, we need to use their local connectivity patterns for identifying meaningful temporal subgraphs in their fabric. Following the approach of \cite{Kovanen_JSTAT2011}, we'll use the concept of temporal adjacency defined above to introduce temporal connectivity and temporal subgraphs.

\begin{definition}\emph{Weak temporal connectivity.} Two events $e_i$ and $e_j$ are temporally weakly connected, if without considering the directionality of adjacency, there is a sequence of temporally adjacent events between them.\label{def:temporal_connectivity}
\end{definition}

\begin{definition}\emph{Weak $\Delta t$-connectivity}. Two events $e_i$ and $e_j$ are weakly $\Delta t$-connected if they are temporally connected through $\Delta t$-adjacent events, without considering the directionality of $\Delta t$-adjacency.
\end{definition}

The above definitions of temporal connectivity are weak in the sense that the directions of adjacency do not matter. Their motivation is to ensure that temporal subgraphs -- as defined below -- are connected both topologically and temporally.

\begin{definition}\emph{Connected temporal subgraphs}.
A connected temporal subgraph consists of a set of events where all pairs of events are weakly temporally connected.\label{def:subgraph}
\end{definition}

Note that in the above definition we have left out the word "weak" because there cannot be strong connectivity between temporal network events (there cannot be any loops in time).

\begin{definition}
\emph{$\Delta t$-connected temporal subgraphs.} A $\Delta t$-connected temporal subgraph consists of a set of events where all pairs of events are weakly $\Delta t$-connected. The subgraph is called \emph{valid} if no events are skipped when constructing the subgraph; \emph{i.e.}, for each node's time span in the subgraph, all events that can be included are included.\label{def:delta_t_subgraph}
\end{definition}

\begin{definition}\emph{Maximal valid connected subgraphs}.
A maximal valid  connected temporal subgraph is a connected temporal subgraph that contains all events that can be added to it such that all its event pairs are temporally connected.\label{def:max_valid_subgraph}
\end{definition}

\begin{definition}\emph{Maximal valid $\Delta t$-connected subgraphs}.
A maximal valid $\Delta t$-connected temporal subgraph is a $\Delta t$-connected temporal subgraph that contains all events that can be added to it such that all its event pairs are $\Delta t$-connected. \label{def:max_valid_deltat_subgraph}
\end{definition}
Note that by definition, maximal valid $\Delta t$-connected subgraphs are themselves subgraphs of maximal valid temporal subgraphs.

\subsection{Definitions: time-respecting path and $\Delta t$-constrained time-respecting path}

As the final building block before discussing weighted event graphs, we will next focus on temporal-network paths that define which nodes can reach one another and when. Similarly to static-network paths that are sequences of nodes joined by edges, the events of temporal networks form paths in time that connect nodes. For a temporal-network path to be meaningful, it has to respect the direction of time:

\begin{definition}\emph{Time-respecting path.}
An alternating sequence of nodes and undirected events $P=[ v_1, e_1(i_1,j_1,t_1),  \dots, e_n(i_n,j_n,t_n), v_{n+1} ] $ is a time-respecting path  
if the events are consecutive in time and each consecutive pair of events is temporally adjacent, $e_k\rightarrow e_{k+1}$ for all $k < n $, and $v_k,v_{k+1} \in \{ i_k,j_k \}$, such that $v_k\neq v_{k+1}$.
 If the events are directed, then additionally each event's target node must be the source node of the next event on the path, $j_k=i_{k+1}$. For notational convenience, we can omit the nodes, defining a time-respecting \emph{event path} $P_e =[ e_1(i_1,j_1,t_1), \dots, e_n(i_n,j_n,t_n) ]$.
\label{def:time_respecting_path}
\end{definition}

For events with zero duration,  $P_e=\left[e_1(i,j,t_1),e_2(j,h,t_2),e_3(h,l,t_3)\right]$ is a time-respecting path if $t_1<t_2<t_3$, in other words, if $\delta t = t_{k+1}-t_k > 0 \; \forall k=1,2,\,e_k\in P_e$. The inequality follows from the requirement that vertices only participate in at most one event at a time.
For events with durations, the next event on the path cannot begin before the first event is finished: $P_e=\left[e_1(i,j,t_1,\tau_1),e_2(j,h,t_2,\tau_2),e_3(h,l,t_3,\tau_3)\right]$ is a time-respecting path when $t_1+\tau_1<t_2$ and $t_2+\tau_2<t_3$. That is, if $\delta t = t_{k+1}-t_k-\tau_k>0 \; \forall k=1,2,\,e_i\in P_e$. Note that time-respecting paths are always directed, no matter whether the events themselves are directed or not, and their direction follows the arrow of time.

Finally, as a special case of time-respecting paths, we define a subset of them where the events have to follow one another within some specific time limit $\Delta t$.

\begin{definition}
\emph{$\Delta t$-constrained time-respecting paths}. A time-respecting path is $\Delta t$-constrained if all its consecutive pairs of events are $\Delta t$-adjacent, \emph{i.e.}, all consecutive events follow one another with a time difference of no more than $\Delta t$:  $\delta t(e_k,e_{k+1})%
\leq\Delta t \,\forall k<n$.
\label{def:delta_t_constrained_trp}
\end{definition}

\subsection{The weighted event graph D}

Armed with the above definitions, our aim is now to map the original temporal network onto a static representation that retains information of the time-respecting paths of the network (Def. \ref{def:time_respecting_path}) as well as the time differences $\delta t$ between events on such paths. For a temporal network $G=(V_G,E_G,T)$, let $A_G = \{ (e_i,e_j) | e_i \rightarrow e_j; \, e_i,e_j \in E_G \} \subset E_G\times E_G$ be the set of all temporal adjacency relations between the events $E_G$ of $G$ (see Def.~\ref{def:temporal_adjacency}). We are now ready to define the weighted event graph:

\begin{definition}\emph{Weighted event graph.} The weighted event graph of a temporal network $G$ is a weighted graph $D=(V_D,L_D,w)$,
where the nodes $V_D=E_G$, links $L_D= A_G$, and the weights of the edges are given by $w(e_i,e_j)= \delta t(e_i,e_j)$.
\end{definition}\label{def:D}

\begin{figure}[t]
\begin{center}
\includegraphics[width=0.9\linewidth]{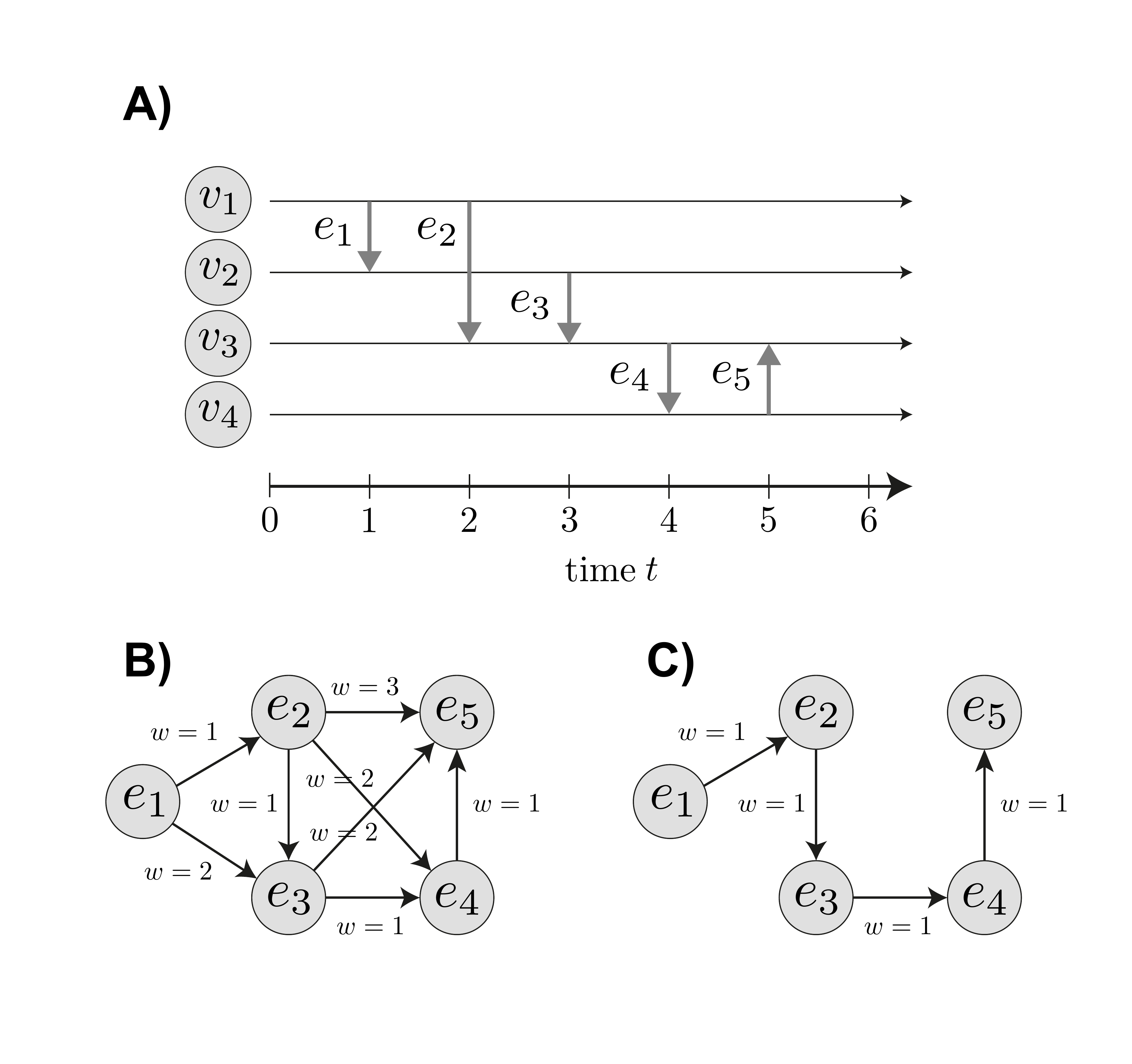}
\caption{Constructing the weighted event graph $D$. Panel A) shows the timeline representation of the original temporal network $G$ with vertices $v_1$, $v_2$, $v_3$, and $v_4$, and events $e_1\ldots e_5$. Panel B) shows the weighted event graph $D$ that corresponds to $G$. Panel C) displays the thresholded version $D_{\Delta t}$ with $\Delta t=1$.} \label{fig:schematic}
\end{center}
\end{figure}

In other words, the weighted event graph $D$ is a directed graph 
whose vertices map to the events of $G$, whose directed links $L_D$ map to the adjacency relations $e \rightarrow e'$ between $G$'s events, and whose link weights $W$ indicate for each adjacency relation the time difference $\delta t$ between the two events. For a schematic example of how $D$ is constructed, see Figure $\ref{fig:schematic}$.

Because of how the weighted event graph $D$ is constructed, it is directed, with the direction of its links following the direction of time. Consequently, because there cannot be any loops in time, it is also acyclic and therefore a DAG (Directed, Acyclic Graph). This provides certain computational advantages.

Note that in this mapping, \emph{isolated} events, that is, single events connecting pairs of nodes that have no other events in $G$,
become isolated zero-degree nodes of the event graph $D$.
It may be convenient to entirely remove such zero-degree nodes from $D$.

\subsection{Constructing the weighted event graph computationally}

The weighted event graph presentation $D$ of an empirical temporal network $G$ can be constructed computationally by inspecting the timeline of the events of each of $G$'s nodes separately. For practical purposes, to save memory, we recommend  setting a maximum value of the time difference between events, $\Delta t_{\max}$, above which two events will not be connected in $D$. Typically, the problem at hand yields a natural time scale, \emph{e.g.} when studying processes of contagion it is not necessarily meaningful to connect events that take place several months apart. However, if memory consumption is not a problem, one can use the entire available time range and set $\Delta t_{\max}=T$, where $T$ is the largest time in the data set.

As the temporal adjacency of two events requires that the events share at least a single temporal-network node, it is convenient to compute the adjacencies around each of these nodes separately. 
For instantaneous events (that is, events that have no durations), one can construct a time-ordered sequence of events containing node $i$: $\{e_{i1}, e_{i2}, \ldots, e_{ik}\}$. Then, it is straightforward to iterate over this sequence: begin at each event $e_{il}$ and scan forward until the time cap $\Delta t_{\max}$ is met, that is, $t_{in}-t_{il}>\Delta t_{\max}$. While scanning, connect each intermediate event $e_{im}$ with the focal event with the weight $w_{il,im}=\delta\left(e_{il},e_{im}\right)=t_{im}-t_{il}$ (unless they are already connected by a previous sweep, which is possible for repeated events between the same pair of nodes). Rather similar but slightly more complicated algorithms can be used for temporal networks with events that have durations or even higher-order events that contain more than two nodes.

Creating these sequences of events and sorting them can be done in $\mathcal{O}(|E_G|\log|E_G|)$ time. Because each step of the algorithms yields one connection in $D$ (note that some links may be visited twice), the total runtime of the algorithm is $\mathcal{O}(|E_G|\log|E_G|+|E_D|)$. However, even though the computation of event graphs is quite economic, this representation can have significantly higher memory complexity than the original temporal network representation. While a temporal network can be represented as an event sequence, which requires  $\mathcal{O}(|E_G|)$ of memory, event graphs can occupy way more memory than such sequences. In the worst case, their memory complexity is $\mathcal{O}(|E_G| a)$, where $a$ is the maximum number of events a node participates in. %

If one is only interested in the connectivity, that is, in knowing whether there is a path between two events regardless of the rest of the paths, then it is possible to use the directed and acyclic nature of the event graph $D$ as an advantage when doing the computations. The directed connectivity in a DAG is a transitive relationship, which means that one can always remove edges whose source and target nodes are connected by some other path without affecting the overall connectivity (weak or directed). Taken to the extreme, this will lead to the \emph{transitive reduction} of $D$. For all edges removed in this way, the weight of the removed edge is smaller than the weights of the edges in the indirect path. This is a useful fact when thresholding the network (as is done in Section \ref{sec:threshold}).

A computationally convenient way of removing some (but not necessarily all) of the transitively redundant edges from $D$ is to simply stop the above described algorithm after the first iteration for each node \cite{Mellor2017}. This trick will only work for weighted temporal event graphs built with undirected adjacency relations, but it will bound the out-degree of the nodes to 2, which can dramatically reduce the time and memory complexity of the network creation algorithms in some cases.

\subsection{Thresholding the weighted event graph}\label{sec:threshold}

A key strength of the weighted event graph approach is that the event graph $D$ can be quickly thresholded so that the resulting graph $D_{\Delta t}$ only contains directed links between events that follow one another within a time $\Delta t$ in the original temporal network $G$. Formally, $D_{\Delta t}$ is defined as follows: let $G=(V_G,E_G,T)$ be a temporal network and $A_{G,\Delta t}\subset E_G\times E_G$ the set of all $\Delta t$-adjacency relations between its events $E_G$. %
We can now define thresholded event graph $\Delta t$:

\begin{definition}
\emph{Thresholded event graph $D_{\Delta t}$}. The thresholded event graph $D_{\Delta t}$ of $G$ is the graph $D_{\Delta t}=(V_{D_{\Delta t}},L_{D_{\Delta t}},w)$ with vertices $V_{D_{\Delta t}}$, directed links $L_{D_{\Delta t}}$, and link weights $W_{D_{\Delta t}}$, so that $V_{D_{\Delta t}}= E_G$, $L_{D_{\Delta t}}= A_{G,\Delta t}$, and $w(e_i,e_j)= \delta t(e_i,e_j)\leq\Delta t$.
\end{definition}

In other words, $D_{\Delta t}$'s nodes are again vertices of $G$, its directed links are $\Delta t$-adjacency relations between the events of $G$, and its link weights are the time differences $\delta t$ between $\Delta t$-adjacent events where by definition  $\delta t\leq \Delta t$. Therefore, $D_{\Delta t}$ is a subgraph of $D$ that only contains links between events that follow one another within $\delta t \leq \Delta t$.

While $D_{\Delta t}$ can in principle be constructed directly from $G$ using $\Delta t$-adjacency relations from the beginning, this is not the fastest approach if  one wants to vary $\Delta t$. Rather, it is much faster to first construct $D$ up to the maximum $\Delta t_{\mathrm{max}}$ and then threshold it to $D_{\Delta t}$ by discarding all links with weights above the  chosen value of $\Delta t$. This is how the thresholded weighted event graph $D_{\Delta t}$ is always constructed in practice.

Typical use cases for the thresholded event graph $D_{\Delta t}$ include setting maximal values of the allowed time difference between events to account for processes such as deterministic SIR (Susceptible-Infectious-Recovered) or SIS (Susceptible-Infectious-Susceptible), defined so that an  infectious node can only infect other nodes through events that take place within a certain constant time since the time of its infection. For modelling certain transport processes, \emph{e.g.} passenger trips through public transport networks or the air transport network, limiting the allowed range of $\delta t$ from both above and below might be appropriate instead of using an upper limit $\Delta t$ only. In such cases, the lower limit would indicate the shortest possible transfer time between vehicles, and the upper limit would correspond to the maximum time that the passengers are willing to wait for their connection.

It is often useful to sweep through a range of allowed values of $\delta t$, as in the temporal-network percolation studies discussed later in this Chapter. When thresholding so that all links of $D$ with weights below the limit $\Delta t$ are retained while varying $\Delta t$, one can obtain huge savings in computational time by the following procedure: 1) order the links of $D$ in increasing order of weight, 2) begin with an empty network, 3) add links one by one, 4) after each link addition, mark down current the threshold value $\Delta t$ and compute the quantities of interest such the sizes of components in the network. Here, one can easily and quickly keep track of component sizes by initially assigning each node to its own component and then always
checking if the newly entered link merges two components or not. In fact, with the help of disjoint-sets forest data structure, one does not even need to actually construct the network: it is enough to keep track of the component mergings and their sizes. 
This procedure is similar to the ones used for analysing connectivity of static networks in percolation studies \cite{Newman2001Fast,Onnela2007}.

\section{How to interpret and use weighted event graphs}

\subsection{How the basic features of $D$ and $D_{\Delta t}$ map onto features of $G$}

Let us begin dissecting the weighted event graphs by mapping out simple correspondences between some features of $D$ and features of $G$. In the following, for the sake of simplicity, we shall consider the original temporal network $G$'s events as undirected and instantenous. Further, we assume that the weighted event graph $D$ has been constructed using the whole available time range, that is, with time differences up to $\Delta t_{\max}=T$. By definition, the thresholded version of the event graph $D_{\Delta t}$ only contains links between $\Delta t$-adjacent events, that is, events with time differences less than $\Delta t$.

First, as already evident, the elements of $D$ map to the elements of $G$ so that the nodes of $D$ are events in $G$, the links of $D$ are temporal adjacency relations between the events of $G$, and the link weights of $D$ indicate the times between adjacent events in $G$. %
The \emph{in- and out-degrees} of a node of $D$ indicate the numbers of \emph{temporal adjacency relations} between the corresponding event of $G$ and previous/future events of the two nodes that the event connects: the in-degree of node $e_i$ of $D$ (event $e_i$ of $G$) is the number of events that took place earlier than $e_i$ and involved either or both of the connected nodes. The out-degree is the number of similar future events.
For $D_{\Delta t}$, the in- and out-degrees of nodes correspond to the numbers of past and future events of the event's endpoint nodes in $G$ within a time $\Delta t$. This latter property could be useful, e.g., for studying temporal threshold models (see, e.g., ~\cite{Karimi2013,Takaguchi2013,Backlund2014}) where the process of contagion is triggered by infection from multiple sources within some short time range.

Due to $D$'s construction, \emph{a directed path in $D$} is a \emph{time-respecting path} in $G$, and vice versa. If we define (without the loss of generality) a vertex path $P_v$ in a graph as a sequence of vertices joined by an edge, then we can formalize this relationship:
\begin{theorem}
\emph{Path equivalence.}
A path $P$ is a vertex path in $D$ if and only if $P$ is a time-respecting event path in $G$.
\end{theorem}
Put in another way, if $\mathcal{P}_v(D)$ is the set of all vertex paths in the graph $D$, and $\mathcal{P}_e(G)$ is the set of all time-respecting event paths in $G$, then $\mathcal{P}_v(D)=\mathcal{P}_e(G)$.

For $D_{\Delta t}$, the corresponding time-respecting path in $G$ is in addition $\Delta t$-constrained and so the time difference between its consecutive events is always less than $\Delta t$ (see Def.~\ref{def:delta_t_constrained_trp}). 
\begin{theorem}
\emph{Constrained path equivalence.}
A path $P$ is a vertex path in $D_{\Delta t}$ if and only if $P$ is a $\Delta t$-constrained time-respecting event path in $G$.
\end{theorem}
Now, if in addition we denote by $\mathcal{P}_e(G,\Delta t)$ the set of all  $\Delta t$-constrained time-respecting event paths in $G$, then $\mathcal{P}_v(D_{\Delta t})=\mathcal{P}_e(G,\Delta t)$.

If the events are instantaneous, then, additionally, the sum of link weights of a path in $D$ equals the \emph{latency} or \emph{temporal distance} of the corresponding path in $G$, \emph{i.e.}, its duration. %
This property can be rather useful: \emph{e.g.} for undirected, instantaneous events, the lowest-weight path from $e_i$ to $e_j$ in $D$ equals the fastest time-respecting path in $G$ from event $e_i$ to event $e_j$ (again, expressing time-respecting paths in terms of events rather than nodes). Hence, it is possible to directly use $D$ for computing centrality measures that are defined in terms of  time-respecting paths or shortest time-respecting paths.

Because of the above, \emph{the set of downstream nodes in $D$} reached by following the directed links of $D$ from its node $e_i$ equals the reachable set of event $e_i$ in $G$, in other words, the set of all events in $G$ that can be reached from $e_i$ through time-respecting paths (its "future light-cone"). Likewise, \emph{the set of upstream nodes} that can be reached by following $D$'s links in reverse direction equals the set of all events in $G$ that can lead to $e_i$ through time-respecting paths: the set of events that may influence $e_i$ ("past light-cone"). For $\Delta t$, the sets of upstream/downstream nodes come with the additional constraint that they must be reachable through $\Delta t$-constrained time-respecting paths.

Finally, the weakly connected components of $D$ (more on components later) correspond by definition to maximal valid temporal subgraphs in $G$ (Definition \ref{def:max_valid_subgraph}); for $D_{\Delta t}$, the weakly connected components correspond to maximal valid $\Delta t$-connected subgraphs (Definition \ref{def:max_valid_deltat_subgraph}).

All the above correspondences are summarized in Table~\ref{tab1} for $D$ and in Table~\ref{tab2} for $D_{\Delta t}$.

\begin{table}
\caption{Correspondence between features of the weighted event graph $D$ and the original temporal graph $G$.}\label{tab1}  

\begin{tabular}{p{4.3cm}p{7cm}}
\hline\noalign{\smallskip}
Feature in $D$ & Feature in $G$  \\
\noalign{\smallskip}\svhline\noalign{\smallskip}
node $V_D$ & event $E_G$\\
link $L_D$ & temporally adjacent pair of events $e_1\rightarrow e_2$\\
link weight $w$ & time difference $\delta t$ between adjacent events \\
in-degree $k_\mathrm{in}$& \# of previous events of the event's endpoint nodes \\
out-degree $k_\mathrm{out}$ & \# of future events of the event's endpoint nodes \\
directed vertex path $P_v$ & time-respecting event path $P_e$\\
sum of weights on path $P$ & duration of time-resp.~path  $P$ (if events instantaneous) \\
set of downstream nodes for $v_D$ & set of events reachable from $e_G$ ("future light-cone")\\
set of upstream nodes for $v_D$ & set of events that can influence event $e_G$ ("past light-cone") \\
weakly connected component & maximal valid temporal subgraph  \\

\noalign{\smallskip}\hline\noalign{\smallskip}
\end{tabular}
\end{table}

\begin{table}
\caption{Correspondence between features of the $\Delta t$-thresholded event graph $D_{\Delta t}$ and the original temporal graph $G$.}\label{tab2}  

\begin{tabular}{p{4.3cm}p{7cm}}
\hline\noalign{\smallskip}
Feature in $D_{\Delta t}$ & Feature in $G$  \\
\noalign{\smallskip}\svhline\noalign{\smallskip}
node $v_D$ & event $e_G$\\
link $l_D$ & $\Delta t$-adjacent pair of events $e_1\xrightarrow{\Delta t} e_2$\\
link weight $w$ & time difference $\delta t$ between adjacent events \\
in-degree $k_\mathrm{in}$& \# of previous events of the event's endpoint nodes within $\Delta t$\\
out-degree $k_\mathrm{out}$& \# of future events of the event's endpoint nodes within $\Delta t$\\
directed vertex path $P_v$ & $\Delta t$-constrained time-respecting path $P_{e}$\\
sum of weights on path $P$ & duration of time-resp.~path $P$ (if events instantaneous) \\
set of downstream nodes for $v_D$ & set of events reachable from $e_G$ through $\Delta t$-constrained time-respecting paths \\
set of upstream nodes for $v_D$ & set of events that can influence event $e_G$ through $\Delta t$-constrained time-respecting paths\\
weakly connected component & maximal valid $\Delta t$-connected subgraph  \\

\noalign{\smallskip}\hline\noalign{\smallskip}
\end{tabular}
\end{table}

\subsection{Temporal motifs and $D$}

The concepts of $\Delta t$-adjacency, $\Delta t$-connectivity and temporal subgraphs are intimately related to \emph{temporal motifs}~\cite{Kovanen_JSTAT2011,Kovanen_book,Kovanen_PNAS}. The concept of \emph{network motifs} was originally introduced for static networks by Milo \emph{et al.}~\cite{Milo_Science2002} in 2002. They defined network motifs as classes of isomorphic induced subgraphs with cardinality higher in the data than in a reference system, usually the configuration model. Milo \emph{et al.} showed that similar networks had similar characteristic network motifs, suggesting that motif statistics are informative of the function of the system and could be used to define universality classes of networks~\cite{Milo_Science2004}.

Similarly to static-network motifs, temporal motifs are one way of looking at frequent, characteristic patterns in networks. In this case, the patterns are defined in terms of both topology and time. For temporal motifs, a natural starting point is to use the definition of $\Delta t$-connected subgraphs (Def.~\ref{def:delta_t_subgraph}), and to look at temporal-network entities where a sequence of interaction events unfolds in the same way. As an example, the sequence where A calls B calls C calls A forms a triangular $\Delta t$-connected subgraph if all calls follow one another with a time difference of no more than $\Delta t$. Note that here we consider the events to be directed, but using undirected events is also possible. 

Such temporal-topological patterns reflect the dynamics of the system in question. Therefore, their characterization can improve our understanding of various complex systems, \emph{e.g.}, of temporal networks whose structure reflects the nature of human social interactions and information processing by groups of people. As an example, Ref.~\cite{Kovanen_PNAS} showed that there is a tendency of similar individuals to participate in temporal communication patterns beyond what would be expected on the basis of their average interaction frequencies or static-network structure, and that the temporal patterns differed between dense and sparse regions of the network. These observations relied on the timings of the communication events, reflected in their $\Delta t$-connectivity.

Temporal motifs are formally defined as equivalence classes of isomorphic $\Delta t$-connected, valid temporal subgraphs (Def.~\ref{def:delta_t_subgraph}), where the isomorphism  takes into account both the topology of the subgraph and the temporal order of events. With this definition, the two-call sequences A calls B calls C and D calls E calls F both belong to the same two-event equivalence class (if the $\Delta t$-adjacency condition is met). 

The temporal-topological isomorphism problem can be solved using a trick that combines the event graph approach presented in this Chapter with the topology of the subgraph in the original network: a "virtual" node is added onto each (event) link, analogous with an event node in $D$. This virtual node is then connected with a directed arrow to the event that immediately follows it~\cite{Kovanen_JSTAT2011,Kovanen_book}; this is a limited version of temporal adjacency, as only the next event is considered. The directed arrows between the virtual nodes determine the order of events in the subgraph. The virtual nodes are then assigned a "color" different than the original nodes, and the isomorphism problem is finally solved using static-network algorithms for directed, coloured graphs, such as Bliss~\cite{bliss}.

The procedure for obtaining temporal-motif statistics from empirical temporal networks with time-stamped events is as follows~\cite{Kovanen_JSTAT2011}, for a given value of $\Delta t$ and a chosen motif size $s$ measured in events: 
\begin{enumerate}[label=\roman*.]
\item Find all maximal $\Delta t$-connected subgraphs $E^\ast_\mathrm{max}$ of $G$.
\item Find all valid temporal subgraphs $E^\ast\subset E^\ast_\mathrm{max}$ of size $s$. 
\item Solve the isomorphism problem to find equivalence classes for all $E^\ast$.
\item Count the number of motif instances in each equivalence class, and compare against a chosen null model.
\end{enumerate}

For details including pseudocode for the required algorithms, we refer the reader to~\cite{Kovanen_JSTAT2011,Kovanen_book}.

Here, if one wants to compare motif statistics for a range of values of $\Delta t$, as is often the case, the weighted event graph approach helps to substantially reduce computational time for step (i) of the above procedure. While it is possible to generate the maximal $\Delta t$-connected subgraphs for each value of $\Delta t$ separately from $G$'s events using brute force, the threshold sweep approach outlined in Section~\ref{sec:threshold} is a much better solution. 

With this approach, one simply needs to generate the weighted event graph $D$ and then threshold it by discarding all edges with weights above each $\Delta t$. If one wants to compute motif statistics for, say $\Delta t_1<\Delta t_2<\ldots<\Delta t_\mathrm{max}$, the fastest way is to begin with an empty network and sort links by increasing weight. Then, one add links up to link weight $\Delta t_1$ and either store $D_{\Delta t_1}$ or compute the quantities of interest, then add more links up to $\Delta t_2$ and do the same, and repeat up to $\Delta t_\mathrm{max}$. Note that here, one does not initially need to construct the whole $D$ which might cause memory problems: building it up to $\delta t = \Delta t_\mathrm{max}$ is sufficient.

\subsection{Components of D and temporal-network percolation}

\subsubsection{Measuring component size}

Let us next discuss the components of $D$ (and $D_{\Delta t}$) in more detail. First, because the event graph $D$ is directed, the usual complications of defining components in directed networks apply. However, because $D$ is also acyclic, there cannot be any strongly connected components, where all nodes are reachable from all other nodes. Therefore, connected components of $D$ ($D_{\Delta t}$) can only be %
weakly connected by definition.

For the purposes of our interest, we can focus on three types of components: (i) maximal \emph{weakly connected} components of $D$, where all nodes of $D$ are joined by a path if the directions of $D$'s links are ignored and no more nodes can be added, corresponding to maximal valid temporal subgraphs of $G$ (Def.~\ref{def:subgraph}); (ii) maximal \emph{out-components}, uniquely defined for each node $v_D$ of $D$, so that all nodes in the out-component can be reached from the focal node $v_D$, and (iii) maximal \emph{in-components}, again defined uniquely for each $v_D$, so that the focal node can be reached from all of the component's nodes. These definitions do not change if we use $D_{\Delta t}$ (however, the thresholded $D_{\Delta t}$ is of course expected to have a different component structure, generally with more components than $D$). In the following, we will for simplicity talk about $D$ only, but everything holds for $D_{\Delta t}$ as well.

Let us next discuss the properties -- in particular, the concept of size -- for components of $D$ defined using any of the above definitions. 

\begin{figure}
\begin{center}
\includegraphics[width=0.9\linewidth]{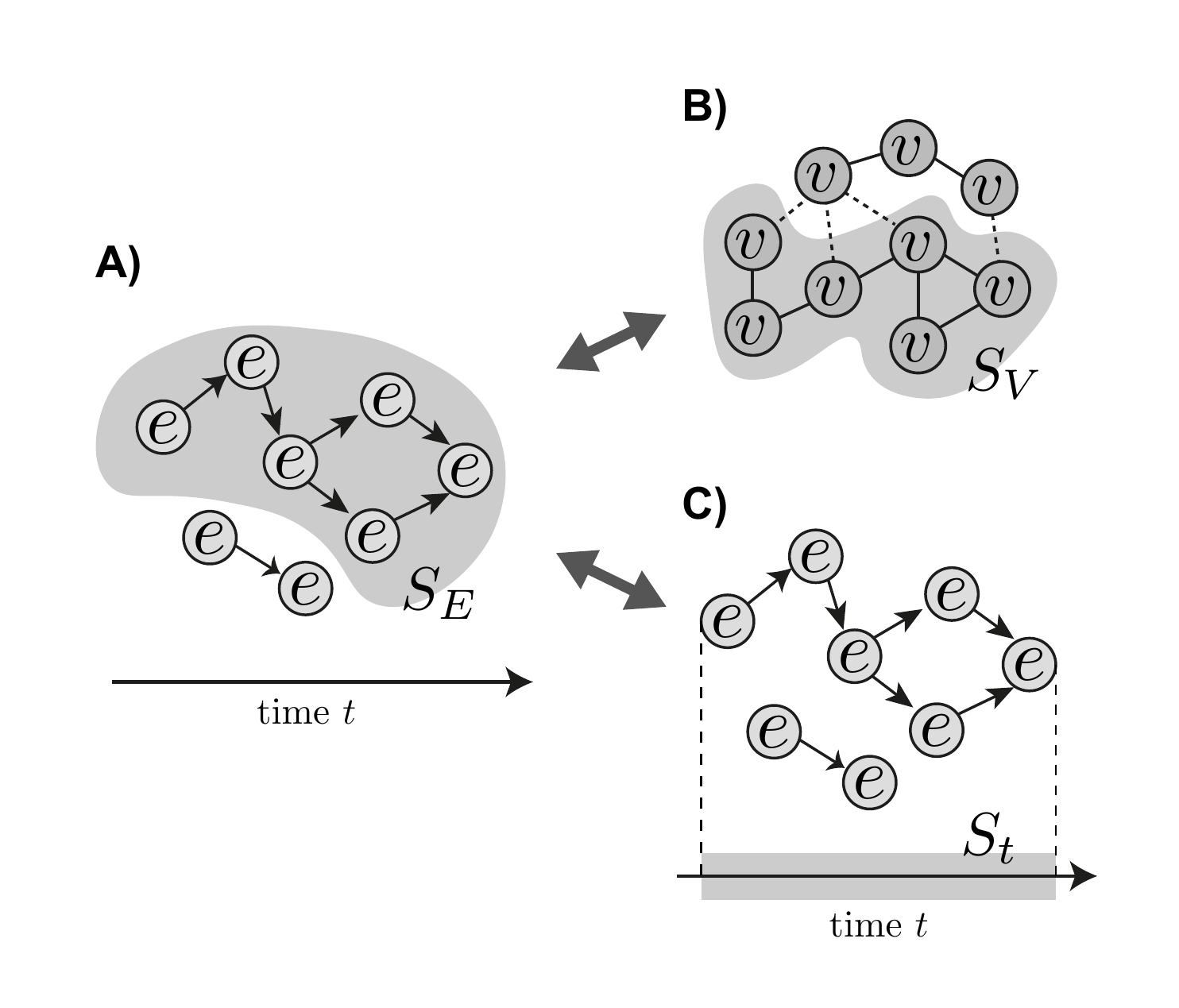}
\caption{Panel A: The shaded area indicates the size $S_E$ of a component of $D$, measured in the number of nodes of $D$ (events of $G$) involved in the component. Panel B: the size $S_V$ of the same component, measured as the number of involved vertices in the original graph $G$ as indicated by the shaded area. Panel C: The third way of measuring component size, the lifetime $S_t$ of the component measured as the time difference between its last and first events.} \label{fig:components}
\end{center}
\end{figure}

First, the most straightforward way to define component size is to count the number $S_E(\mathbb{C})$ of the event graph $D$'s nodes that belong to the component $\mathbb{C}$. This is equal to the number of events in the original temporal network $G$ that belong to the same component, and $S_E(\mathbb{C}) \in [0,|E_G|]$. For a schematic illustration, see Fig.~\ref{fig:components}, panel A.

Second, one can map the nodes of $D$ in component $\mathbb{C}$ back to the events of the temporal network $G$ and count the number of vertices involved in the events, $S_V(\mathbb{C}) \in [0,|V_G|]$. This is the "spacelike" definition of size (see Fig.~\ref{fig:components}, panel B).

Third, because the event nodes in $D$ come with time stamps -- the events take place at specified times -- one may think of a "timelike" size: the \emph{duration} (that is, the lifetime) of the component $S_t(\mathbb{C}) \in [0,T]$, the time difference between $\mathbb{C}$'s last and first event. This is illustrated in Fig.~\ref{fig:components}, panel C.

Note that these measures of size may or may not be correlated in a temporal network. In a random, Erd\H{o}s-R\'enyi-like temporal network they on average are (see Ref.~\cite{Kivela2018}). In this case, one can think of a single "giant" temporal component that encompasses most of the events in $D$ and nodes in $G$ and that lives for the entire observation period of the temporal network. However, this is a special case, and one can equally well think of networks where the different types of "giant" components are separated. As an example, there can be a short-lived, "spacelike" component that spans most of the nodes in $G$ but contains only a small fraction of the nodes in $D$ because of its short lifetime. There may also be several such components during the lifetime of the network. Further, one can also envision a persistent component that spans the whole time range but involves only a small number nodes that repeatedly and frequently interact: this component is large in $S_t$ but vanishingly small in $S_E$ and $S_V$. Again, multiple such components may coexist.

\subsubsection{Temporal-network percolation analysis with $D_{\Delta t}$} \label{sec:percolation}

When the event graph $D$ is thresholded, its component structure depends on the threshold weight. If the criterion for retaining $D$'s edges is that their weight is below some chosen value of the time difference $\Delta t$, then this parameter can be viewed as the \emph{control parameter} of a percolation problem. The value of the control parameter $\Delta t$ determines the event graph's component structure, in particular its largest component, similarly to the edge weight threshold used in percolation studies on static, weighted  networks (see, e.g., ~\cite{Onnela2007}).

In network percolation, there is a critical value of the control parameter that separates the connected and disconnected phases of the network. When the control parameter reaches this value, connectivity suddenly emerges, reflected in the emergence of a giant connected component that spans a finite fraction of the network. This is measured either as the fraction of nodes or links included in the largest component; whichever measure is used, it is called the the \emph{order parameter} of the percolation problem. 

Here, since the control parameter $\Delta t$ operates on the event graph $D$, the most obvious choice for the order parameter would be the relative size of $D$'s largest (weakly connected) component. As discussed above, its size can be measured as the number $S_E$ of its constituent event nodes in $D$, so that the corresponding order parameter
\begin{equation}
\rho_E(\Delta t)=\frac{1}{|E|}\max S_E,
\end{equation}
where $|E|$ is the number of (event) nodes in $D$ and we've made the dependence on $\Delta t$ explicit. As such, this definition works in a straightforward way and $\rho_E(\Delta t)$ can be expected to behave as a typical order parameter would. 

The size of the components other than the largest component is often used in percolation studies to detect the critical point. Using the above definition of size $S_E$, one can define the \emph{susceptibility}
\begin{equation}
    \chi_E=\frac{1}{|E|}\sum_{S_E<\max S_E}n_{S_E}S_E^2, 
\end{equation}
where $n_{S_E}$ is the number of components of size $S_E$ and the sum is over all components except the largest. The susceptibility diverges at the critical point that separates the connected and fragmented phase, as small components are absorbed into the emerging giant component.

However, as discussed above, one can measure the size of a component of $D$ in two other ways. The "spacelike" way is to count the number of nodes $S_V$ of the original network $G$ that are involved in the component through $D$'s event nodes. Using this definition of component size, we arrive at the order parameter that measures what fraction of $G$ is associated with $D$'s component:
\begin{equation}
    \rho_V(\Delta t)=\frac{1}{|V|}\max S_V,
\end{equation}
where $|V|$ is the number of nodes in $G$. For this order parameter, while one could na\"{\i}vely define the corresponding susceptibility-like measure as
\begin{equation}
    \chi_V=\frac{1}{|V|}\sum_{S_V<\max S_V} n_{S_V}S_V^2 \,,\label{eq:V_susc}
\end{equation}
which may behave in hard-to-predict ways at the critical point -- if something that can be called a critical point even exists. This is because a node $v\in V_G$ that participates in multiple connected events will appear multiple times in the corresponding component of $D$. The nodes of the original network $G$ may, for similar reasons, also belong to multiple components occurring at different times. In other words, the sum $\sum_{S_V<\max S_V} n_{S_V}$ is not a conserved quantity.

The third component size definition captures the time length of components in $D$, leading to an order parameter:
\begin{equation}
    \rho_t(\Delta t)=\frac{1}{T}\max S_t\,,\label{eq:t_order}
\end{equation}
where $T$ is the observation period, \emph{i.e.}, the lifetime of $G$. While one could in principle again define a susceptibility-like measure for this control parameter, as in Eq.~(\ref{eq:V_susc}), this measure would not be too useful. This is because  
multiple components of $D$ can easily co-exist, overlapping in time, and there can be a number of long-lived simultaneous components.

When interpreting the results of percolation studies using weighted event graphs, one should bear in mind that the components of $D$ are weakly connected, that is, all pairs of nodes in the component are connected through a path if the directions of the links are discarded. This means that, in the context of spreading processes, the interpretation of component structure and percolation points is that the component size is an upper bound for the number of nodes that can be infected by the process if it begins inside the component. So for processes constrained from above so that the spreading agent has to move forward from a node within $\delta t$, the observed critical $\Delta t$ is a lower bound: one can only be certain that the spreading process would not percolate below this threshold. 

\subsubsection{Temporal-network percolation: empirical examples}

To illustrate the behaviour of the order parameter and the susceptibility as a function of the event-graph threshold $\Delta t$, we'll next recap some of the results originally published in \cite{Kivela2018}, obtained for three data sets: a large dataset of time-stamped mobile-telephone calls~\cite{Karsai2011}, a dataset on sexual interactions from a study of prostitution~\cite{Rocha2011}, and an air transportation network in the US~\cite{transportbureau}. See~\cite{Kivela2018} for more details on the datasets.

\begin{figure}[t]
\begin{center}
\includegraphics[width=0.999\linewidth]{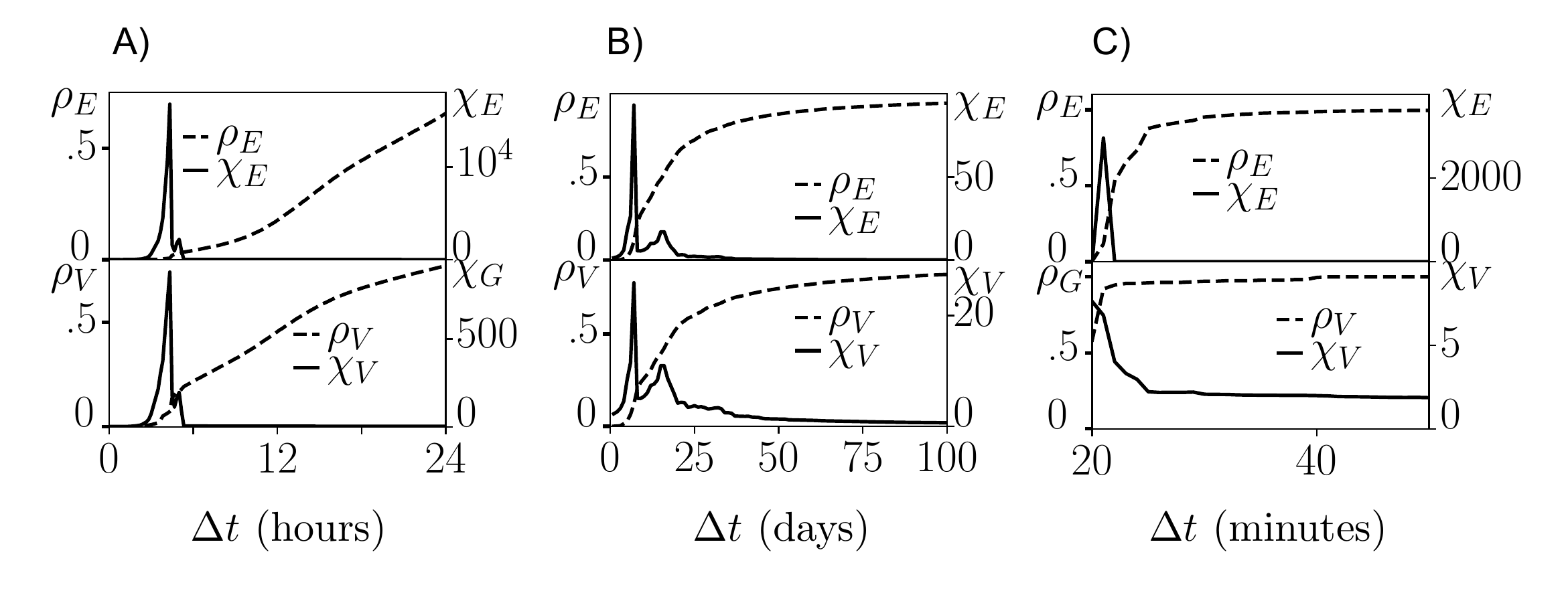}
\caption{The behaviour of the relative largest component size $\rho$ and the susceptibility $\chi$ as a function of $\Delta t$, for three data sets and two variants of the measures. $\rho_E$ and $\chi_E$ are for component size measured in event-graph nodes and $\rho_V$ and $\chi_V$ for size measured in the number of temporal-network nodes involved in the component's events. Panel A: mobile telephone calls, displaying a critical point at around 4 h 20 min. Panel B: sexual interactions, with a critical point at around 7 days (followed by a second peak at $\sim16$ days). Panel C: US air transport, with a critical point at $\sim20$ min. Figure adapted from the original in~\cite{Kivela2018}.} \label{fig:empirical}
\end{center}
\end{figure}

Two versions of relative largest component size (order parameter) and the susceptibility are shown for all datasets in Fig.~\ref{fig:empirical}. The first version is based on the event-graph component size $S_E$ and the second on the number of involved vertices in the original graph, $S_V$. For these data sets, the critical points indicated by the diverging susceptibility are fairly similar for both measures, with the exception of small difference for the air transport network where $\chi_V$ peaks slightly earlier than $\chi_E$. For these datasets, the "timelike" order parameter of Eq.~\ref{eq:t_order} (not shown) does not produce meaningful results; it does not behave like an order parameter for reasons discussed in Section~\ref{sec:percolation}.

The identified critical points are related to characteristic time scales in the systems in question; as an example, they indicate how long a spreading process would typically need to survive in order to eventually reach most of the network. In the case of mobile communication networks, if we imagine, \emph{e.g.} a rumour spreading through the phone calls, the cascade will die out unless the rumour is still relevant and worth spreading for each node after $4$hours and $20$ minutes have passed since the node received it. For the sexual contact network, a sexually transmitted disease can become an epidemic and spread through the network if it remains infectious for longer than $7$ days since being infected. For the air transport network, the identified characteristic time of approximately $20$ minutes is related to the synchronization of connecting flights at airports.

\section{Discussion and conclusions}

Because temporal networks carry information of the times of the interactions between the network's nodes, they allow for the detection of patterns that would be lost were the networks aggregated into static structures. This has led to increased understanding of the dynamics of network structures and processes that unfold on top of networks. The downside of this framework is that it naturally complicates network analysis, because of the additional degrees of freedom brought by the temporal dimension. Temporal networks are, in a way, mixtures of graphs and time series: therefore, if one is not satisfied with studying one of these aspects only, entirely new ways of looking at their structure are required.

In this Chapter, we have presented an approach that projects important features of temporal network dynamics into a static line graph structure: the weighted event graph. Weighted event graphs can be used both to understand the structural features of the temporal networks they encode, as well as to investigate dynamical processes taking place on temporal networks. The weighted event graph framework maps temporal-topological structures onto weighted, directed, acyclic graphs. This is an information-lossless representation of temporal networks, which preserves the time-respecting paths of the original network as well as the timing differences between consecutive events on those paths. Weighted event graphs are particularly useful for studying paths, structures, and processes where one wants to set constraints to the times between successive events ($\Delta t$-connectivity), in other words, where the events have to follow one another quickly enough. 

Beyond the examples discussed in this Chapter (temporal motifs, temporal-network percolation), one can envision many uses for weighted event graphs. In theory, any method or approach, which has been developed around the concepts of paths or walks could benefit from being viewed as a topological problem in weighted event graphs instead of a dynamical problem in temporal networks. Looking beyond the surface, it is clear that many important topics and measures in network science are at least partly based on the path structure of networks, including several approaches in dynamic models on networks, community detection, and centrality measures. As is evident from the cases of percolation analysis and temporal motifs, weighted temporal event graphs can be  useful for both defining understandable concepts and measures as well as providing access to computationally efficient methods for solving temporal-network problems.

There is one rather obvious use of weighted event graphs that we have not discussed yet: the issue of \emph{centrality measures}. The computation of various temporal-network centralities should greatly benefit from weighted event graphs, as they encapsulate the whole set of time-respecting paths (or their $\Delta t$-constrained subset).
Such centralities could straightforwardly be computed using definitions and algorithms developed for static networks but in this case, applied to the event graphs instead. As a bonus, because of the event graph's construction, these measures would be computed for $\emph{events}$ of the original network instead of its vertices. It can be argued that this is -- at least in some cases -- more meaningful than computing quantities for the nodes. Any centrality measure for a node should come with the additional constraint on its valid time range: \emph{e.g.}, because time-respecting paths constantly change, should the  "temporal betweenness centrality" of a node be a quantity that characterizes the node's properties over some time range (up to the entire range of observation of the temporal network), or at some specific point in time, building on the paths that pass through the node at that point? However, with events, the definition is more straightforward: temporal betweenness centrality should depend on the number of (fastest) temporal paths passing through the event. Therefore, at least for instantaneous events, it can be directly and simply calculated from the event graph's directed path structure. 

\begin{acknowledgement}

J.S. acknowledges support from the Academy of Finland, project "Digital Daily Rhythms" (project n:o 297195). M.K. acknowledges support from the Aalto Science Institute and the SoSweet ANR project (ANR- 15-CE38-0011-01).
\end{acknowledgement}

\end{document}